\documentclass[%
 reprint,
superscriptaddress,
 amsmath,amssymb,
 aps,
]{revtex4-2}

\usepackage{graphicx}
\usepackage{dcolumn}
\usepackage{bm}
\usepackage{algorithm}
\usepackage{algpseudocode}
\usepackage[columnwise]{lineno}

\begin{document}

\preprint{APS/123-QED}

\title{Robust Quantum Control using Reinforcement Learning from Demonstration}






\author{Shengyong Li}
\affiliation{Department of Automation, Tsinghua University, Beijing 100084, China}

\author{Yidian Fan}%
\affiliation{Department of Automation, Tsinghua University, Beijing 100084, China}
 
\author{Xiang Li}
\affiliation{Institute of Physics, Chinese Academy of Sciences, Beijing 100190, China}
 
\author{Xinhui Ruan}%
\affiliation{Department of Automation, Tsinghua University, Beijing 100084, China}

\author{Qianchuan Zhao}%
\affiliation{Department of Automation, Tsinghua University, Beijing 100084, China}

\author{Zhihui Peng}
\affiliation{Department of Physics and Synergetic Innovation Center for Quantum Effects and Applications, Hunan Normal University, Changsha 410081, China}

\author{Re-Bing Wu}%
\email{rbwu@tsinghua.edu.cn}
\affiliation{Department of Automation, Tsinghua University, Beijing 100084, China}

\author{Jing Zhang}%
\email{zhangjing2022@xjtu.edu.cn}
\affiliation{School of Automation Science and Engineering, Xi’an Jiaotong University, Xi’an, 710049, China}
\affiliation{MOE Key Lab for Intelligent Networks and Network Security, Xi’an Jiaotong University, Xi’an 710049, China}

\author{Pengtao Song}%
\email{ptsong@xjtu.edu.cn}
\affiliation{School of Automation Science and Engineering, Xi’an Jiaotong University, Xi’an, 710049, China}

\date{\today}

\begin{abstract}
Quantum control requires high-precision and robust control pulses to ensure optimal system performance. However, control sequences generated with a system model may suffer from model bias, leading to low fidelity. While model-free reinforcement learning (RL) methods have been developed to avoid such biases, training an RL agent from scratch can be time-consuming, often taking hours to gather enough samples for convergence. This challenge has hindered the broad application of RL techniques to larger and more complex quantum control issues, limiting their adaptability.
In this work, we use Reinforcement Learning from Demonstration (RLfD) to leverage the control sequences generated with system models and further optimize them with RL to avoid model bias. 
By avoiding learning from scratch and starting with reasonable control pulse shapes, this approach can increase sample efficiency by reducing the number of samples, which can significantly reduce the training time. Thus, this method can effectively handle pulse shapes that are discretized into more than 1000 pieces without compromising final fidelity.  We have simulated the preparation of several high-fidelity non-classical states using the RLfD method. We also find that the training process is more stable when using RLfD. In addition, this method is suitable for fast gate calibration using reinforcement learning. 

\end{abstract}

\maketitle


\section{\label{sec:introduction}Introduction}

In platforms of quantum information, such as superconducting quantum circuits,  Nitrogen-Vacancy (NV) centers and quantum dots, radio frequency (RF) pulses are used to control qubits~\cite{krantz2019quantum, chatterjee2021semiconductor, dobrovitski2013quantum, song2024coherent,song2024experimental}. Shaped laser pulses are also used in trapped ions and  Rydberg atoms~\cite{levine2018high,bruzewicz2019trapped}. 
Many numerical methods have been proposed to design control pulses, such as Gradient Ascent Pulse Engineering (GRAPE), the Krotov method, the Chopped Random Basis (CRAB) method, etc~\cite{khaneja2005optimal, de2011second, goerz2019krotov, caneva2011chopped}. These methods can leverage the system model to directly generate control pulses. 
Furthermore, some analytical methods are developed to generate robust control pulse sequences~\cite{van2017robust, zeng2018fastest}.  Those methods rely on specific cases, require carefully calibrated system parameters, and can generate robust and efficient control pulses.

However, the inaccuracy of system parameters will hinder these methods from achieving high-fidelity controls. Currently, the main obstacles to high-fidelity quantum control in practice include the distortion of the control pulse, the inaccurate parameter determination and calibration, decoherence, crosstalks, non-Markovian properties, and noise from the environment~\cite{guo2024universal,ziman2005all, tripathi2022suppression}. Typically, the above imperfections of the experimental system are not considered in the system model such as Hamiltonian. 
These imperfections prevent quantum control pulses derived by the system model from achieving high fidelity.

Reinforcement learning (RL), a sub-field of machine learning, has been proposed to address the problems discussed above.~\cite{shakya2023reinforcement, jia2020review} . RL has been used to solve complex control problems ~\cite{silver2017mastering, arulkumaran2019alphastar, kiran2021deep}. Reinforcement learning can transform control problems into optimization problems, making it particularly suitable for enhancing the fidelity of quantum control.
Reinforcement learning utilizes direct interactions with quantum systems, thereby having the ability to avoid model bias. Model-free RL has been widely used in quantum control and related tasks~\cite{sivak2022model, sivak2023real, ding2023high, chatterjee2024demonstration, PhysRevApplied.23.024019, PhysRevA.104.053707}. However, model-free RL methods need to interact intensively with quantum systems to collect data to learn a good control sequence, which will impede the generalization of reinforcement learning methods to larger and more complex tasks. Another main difficulty for RL in solving quantum control problems is the high dimensionality of the control field action space. For a quantum control task, there are always multiple control Hamiltonians, and each Hamiltonian requires over hundreds of parameters to describe its pulse characteristics~\cite{heeres_implementing_2017}.  This result is in a high-dimensional parameter space. Random exploration of these high-dimensional spaces is exceptionally challenging, especially when learning from scratch. Model-based RL attempts to learn a system model and then leverages this model to generate additional data, thereby reducing the demand for samples from real quantum systems~\cite{khalid2023sample}.  Whereas, as the dimension of quantum systems increases, learning a precise model of such systems tends to become a very challenging task.


Reinforcement learning from demonstration~\cite{gao2018reinforcement} (RLfD) is a method that combines reinforcement learning and imitation learning. Imitation learning is used to train agents to mimic human experts~\cite{hussein2017imitation}, while reinforcement learning is based on interactions with the environment to facilitate learning. RLfD combines these two approaches, utilizing both demonstration data and environmental interactions to improve learning efficiency. RLfD is used to accelerate the training process by avoiding useless exploration in the early stages of training. Moreover, RLfD is particularly suitable for complex tasks that are difficult to learn from scratch using conventional reinforcement learning methods alone.  In classic RL, demonstration data is always given by human experts or higher level AI systems~\cite{ravichandar2020recent}. Demonstrations can help elucidate key strategies in the task, especially when the action space is large or the state space is complex. In the field of quantum control, demonstrations can be acquired through quantum system models, such as the system Hamiltonian, or through pre-trained RL agents with fewer parameters. It makes RLfD perfectly suitable for quantum control tasks. We perform a detailed research and simulation about using RLfD as a quantum optimal control method. 

Previous model-free RL methods for quantum control always use parameterized gates or slightly discretized pulse shapes, which is not robust to pulse shape distortion. With RLfD, we can discretize the control pulses into piecewise constant pulse with more pieces, which will be more robust to pulse shape distortion.

This work contains four sections. Section 1 introduces basic background information about reinforcement learning in quantum control. Section 2 describes the methods for applying RLfD to quantum control. In Section 3, we present simulation results to support our argument. In Section 4, we conclude our work and discuss about the future work.

\section{\label{sec:methods}REINFORCEMENT LEARNING FOR QUANTUM CONTROL }
\subsection{Reinforcement learning and pulse generation}

Reinforcement learning based on Markov Decision Processes (MDPs) is currently applied to many quantum control problems, both in simulations and experiments~\cite{puterman1990markov}. 

MDP typically consists of six components:
\begin{itemize}
    \item State: State $S$ refers to the set of parameters that fully determine the environment. The current state alone describes the environment without the need for past states, due to the Markovian properties. In quantum control, the state is typically represented by the quantum state vector or density matrix.
    \item Observation: Observation $O$ refers to information received about the state. In quantum control, observations refer to the measurement results obtained from performing positive operator-valued measures (POVMs).
    \item Action: Action $A$ here represents the control pulses applied to the quantum system, causing the environment to transition from one state to the next. It can also be gate parameters or other parameters that impact state transfers.
    \item Reward: Reward $R$  serves as the criterion for evaluating the 'goodness' of an action within a specific state. It provides feedback on the effectiveness of the action in achieving the desired outcome. Here, the reward is also a function of the measurement results. In quantum control, reward is obtained by performing POVMs, thus it's probabilistic.
\end{itemize}

There are also some quantum properties such that single-shot measurements can not extract the whole information from the quantum system and the probabilistic collapse to eigenstates.
In these situations,  MDP describing quantum control is modeled as Partially Observable Markov Decision Processes (POMDPs)~\cite{monahan1982state}. 

The agent learns by maximizing the expected reward.
\begin{equation}
    \theta^* = \arg\max_\theta E(R).
\end{equation}
Here, $\theta$ represents the parameters that parameterize the policy $\pi(a|s;\theta)$.

Due to the physical features of quantum systems, any strong observation of a quantum system will cause a system state collapse, which can be interpreted as an uncontrollable state change. The observation of a quantum state will lead to a probabilistic collapse to an eigenstate. This probabilistic feature leads to a situation where the quantum state cannot be determined with few observations. 

Because of these limitations, in non-feedback quantum control tasks, usually there are no observations and rewards cannot be obtained until the final measurement. In this scenario, every $s, s'$ in $(s, a, r, s')$ is only a placeholder for time~\cite{sivak2022model}. 
For tasks with quantum measurement feedback, the history observations $o_1, \cdots, o_t$ will constitute an estimation of the current state. $o_1, \dots, o_t$ and $ a_1, \dots,a_t$ can be used to estimate the most likely quantum state and then using policy $\pi(a|s)$ to select the best action. 

\begin{figure*}
    \includegraphics[scale=1]{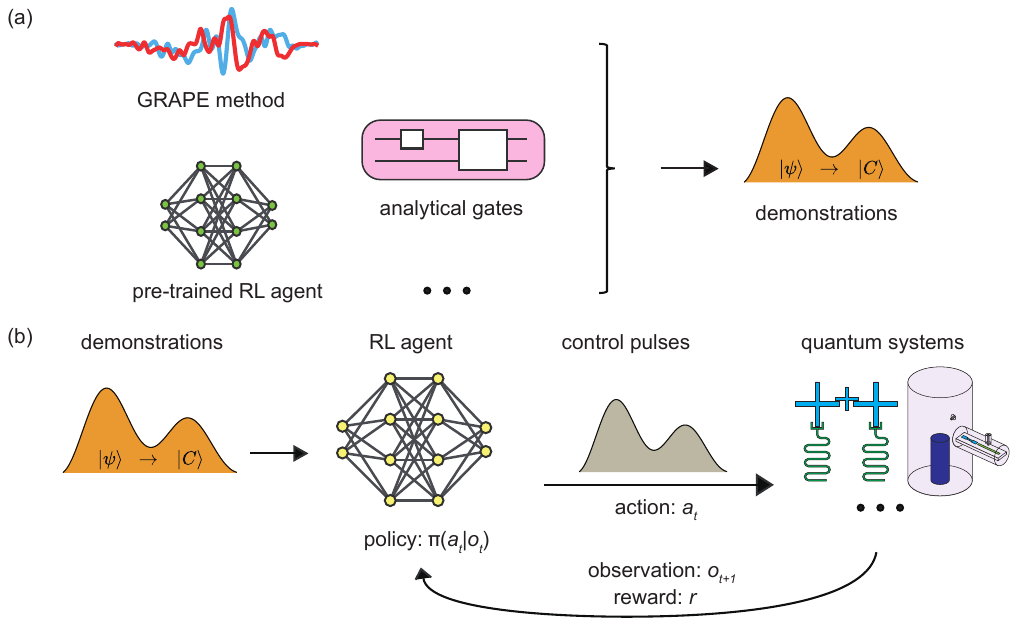}
    \caption{The protocol for practical reinforcement learning from demonstration in circuit QED. (a) Typical ways to get demonstrations for control pulses. GRAPE method and other open-loop optimizing methods are suitable to generate control pulses with system Hamiltonian.
    Pre-trained RL-agent and physical calibrated gates based on previous experiments can also be used to generate demonstration pulses for further optimization. (b) Flow diagram of RLfD for quantum control tasks. The RL agent operates with a policy distribution $\pi(a_t|s_t)$, utilizing the current estimated state to generate an action distribution. In quantum control tasks, actions are control pulses applied to quantum systems. Once the systems are driven by these pulses, measurements are taken to obtain observations  $o_t$ and a reward $r_t$, which assess the effectiveness of the actions applied. $(s_t, a_t, r_t, s_{t+1})$ is used to train the RL agent. In RLfD, demonstration data $(s'_t,  a'_t)$ is also utilized for training.}
    \label{fig:1}
\end{figure*}

Demonstration here means state-action pair $(s, a)$ generated by pre-trained models or human experts. Physical calibrated gates can also be used as demonstrations. Also, one can train a model by optimizing gate parameters rather than pulse shapes. This approach of training gate parameters has fewer parameters, and it’s easier for an RL agent to learn from scratch. Pulse shapes can also be generated using that pre-trained model. Pulses generated can be fed into an RL agent which optimizes pulse shapes to achieve higher fidelity, as shown in Fig. \ref{fig:1}(a) and (b). 
RL methods can also be integrated with simulations. Although accessing the quantum state during control is not feasible in experiments. In simulations, the quantum state can be utilized to aid the training process. This pre-trained RL agent can also be used in experimental quantum control tasks.

One way to utilize the system Hamiltonian is using gradient-based open-loop methods such as GRAPE.
The GRAPE algorithm calculates the gradient of the objective function and the control waveforms. For a system described by the Schrödinger equation, it is assumed that the system is described by the Schrödinger equation in the following format:
\begin{equation}i{\hbar}\frac\partial{\partial t}|\psi(t)\rangle = {H}|\psi(t)\rangle,
\end{equation}
\begin{equation}H = {H}_0+\sum_{k=1}^mu_k[j]H_k.
\end{equation}
The objective function is defined as:
\begin{equation}
J(\vert{\psi(T)}\rangle) = \langle C\vert\psi(T)\rangle\langle \psi(T)\vert C\rangle,
\end{equation}
where$\vert C\rangle$ is the target state, $H_0$ is drift Hamiltonian, $H_k$ is control Hamiltonian, $u_k[j]$ is the discrete control wave, $\vert{\psi(t)}\rangle$ is the state of the quantum system, $T$ is the end time of the control pulses. Therefore, the gradient can be written as 
\begin{equation}
\frac{d J}{du_k[j]} = \frac{d J}{d |\psi(T)\rangle}\frac{d |\psi(T)\rangle}{du_k[j]} + \rm{c.c.}
\end{equation}
In matrix form, it can be abbreviated to $\frac{d J}{du}$.
The control pulse $u$ is updated by $u \leftarrow u + \alpha \frac{d J}{du}$. When $\alpha$ is small enough, the update formula can ensure the increment of objective function $J$~\cite{hindi2004tutorial}.

\subsection{\label{sec:level2}Reinforcement learning from demonstrations}

Learning from demonstration is based on a fact that demonstrations are not optimal but it’s close to the optimal solution. Imitating demonstrations helps the RL agent learn a good action quickly. Exploration and exploitation part can help the agent to explore the environment deeply and then approach the optimal solution. Fig. \ref{fig:1}(a) shows the basic paradigm of utilizing demonstrations for control pulse generation.

Using the GRAPE method we can easily get the state-action pair $(s, a)$, then $(s, a, r, s')$ can be obtained by interacting with the environment. 

Here, we demonstrate pros and cons of the off-policy algorithm, soft actot critic (SAC)~\cite{haarnoja2018soft}, and on-policy algorithm, proximal policy optimization (PPO)~\cite{schulman2017proximal}, for quantum control problems. 

PPO algorithm is an on-policy algorithm~\cite{chatzilygeroudis2019survey} which means the trajectory $(s, a, r, s')$ used in the policy update needs to be collected by the current policy. PPO exploits on-policy trajectories to ensure that the updated policy would not be too far from the previous policy, thereby avoiding significant performance degradation during the training process. It is difficult to integrate demonstrations into PPO algorithm by adding demonstration data to the training data because it breaks its on-policy guarantee, since demonstrations are not sampled by the current agent policy and have different distributions. 

Off-policy algorithms~\cite{prudencio2023survey} like SAC use $Q-$function to help policy to converge, but $Q-$function is difficult to pre-train because we usually do not have enough $(s, a, r)$ demonstrations to make the critic model converge. In this situation, pre-trained policy networks are affected by untrained critical networks and forget learned information. Utilizing only the pre-training policy network is not sufficient.

Since PPO uses $V-$functions rather than $Q-$functions, pre-training is a feasible solution. The $V-$function only needs $(s, r)$ and it is easy to be pre-trained. Policy network and critic network can both be pre-trained, and the pre-trained model can be directly used for the subsequent procedures. For off-policy algorithm, we use demonstration data replay and behavior cloning to make them converge.

Generally, off-policy algorithms like SAC offer better sample efficiency, making them suitable for problems where obtaining data from the environment is challenging. On-policy algorithms like PPO have lower sample efficiency but perform better in training stability and convergence.
In the field of quantum control, problems often have the following characteristics:

\begin{enumerate}
    \item The observation space can be relatively small because the system is not fully observable and the number of reachable states is limited. In non-feedback tasks, observations are just placeholder for time. In measurement-feedback tasks, the observation space is constructed with historical observations.
    \item The dimension of the action space can be relatively high because the control parameters can be quite numerous since the control wave pulse can be divided into many pieces.
    \item Due to the strict timing constraints, the length of each episode is fixed. 
\end{enumerate}

Parameters of action in quantum control tasks always have strong interdependencies. This means that adjusting one control parameter can significantly influence others due to the complex dynamics of quantum systems. 

Learning a good $Q$-function can help the policy network explore the environment more effectively. It understands how different control parameters interact and collectively impact the the evolution of quantum state. This comprehensive understanding allows the agent to predict which kinds of actions will result in better rewards, even if these actions have not been explicitly encountered during training. 

While PPO uses the $V-$function to calculate advantages. It does little to assist exploration. PPO uses random sampling to explore the environment, which is considered less efficient in these contexts. Advantage estimation and gradient clipping help to lower the variance in the training process, which means PPO can provide performance improvement guarantees to a certain degree.  

Algorithms with RLfD are listed in the appendix section \ref{app:sec:sac} and \ref{app:sec:ppo}, as shown in Algorithm \ref{alg:sac} and Algorithm \ref{alg:ppo}. The basic information about these algorithms are also detailed described in appendix.

In learning from demonstration, the difference in data distribution between the demonstrations and the experiments can hinder the agent's learning speed. Therefore, we use demonstration replay to ensure that the agent continuously accesses data from demonstrations, and we use a behavior cloning loss to constrain the agent to explore near the demonstrations.

LayerNorm layers are added into critic network to keep gradient from disappearing and stabilizing the training process~\cite{yue2024understanding, ball2023efficient}.

As for the PPO method, we show how to pre-train a PPO agent with model-generated data.

\section{\label{sec:results}APPLICATIONS}
We demonstrate the advantages of RLfD by applying it to typical quantum control tasks. We first test our method on a simple state preparation task and then extend it to a more complex problem. We get the demonstrations from ideal Hamiltonian model, then add a model bias on Hamiltonian or on the analytical gates to simulate the biases of physical systems.   

    \begin{figure*}
    \centering
    \includegraphics[width=0.96\linewidth]{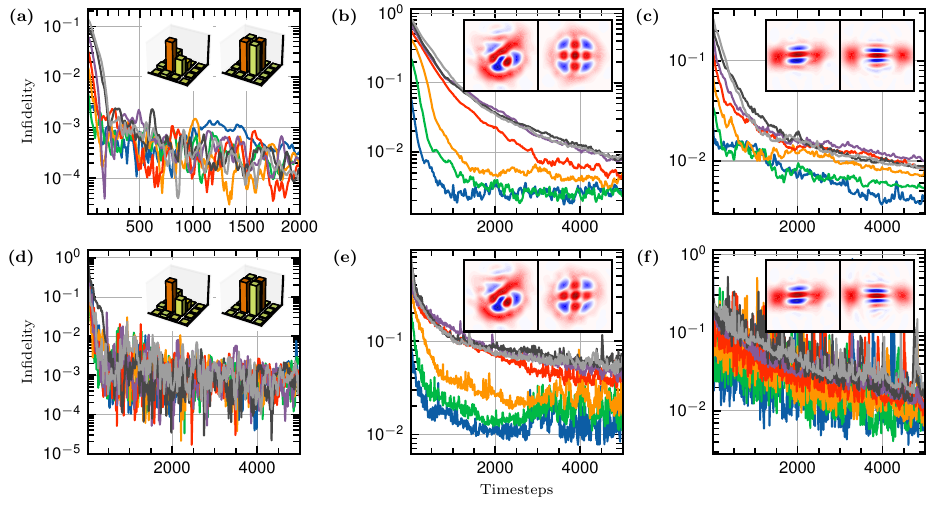}
    \caption{Quantum states preparation with RLfD method by GRAPE pulses. Infidelity versus timesteps for target quantum-state generation under different Hamiltonian error levels (0\%, 5\%, 10\%, 15\%, 20\%, 25\%, 30\% for blue, green, orange, red, purple, brown, gray solid lines respectively). (a), (b), and (c) show the results obtained via the PPO algorithm for generating Bell, binomial, and cat states, respectively. (d), (e), and (f) show the corresponding results via the SAC algorithm. All initial pulse shapes are generated using the GRAPE method. Different colored curves in each sub-figure represent varying degrees of Hamiltonian error. The insets show the density matrix histogram (for Bell state preparation) or Wigner function (for binomial state and cat state) of the state before training (left) and at the end of training (right) with 25\% Hamiltonian error.}
    \label{fig:2}
\end{figure*}


\subsection{\label{sec:bell_state}Bell state preparation}
We initially tested our method on the preparation of a two-qubit Bell state. The Hamiltonian for two qubits coupled via a tunable coupler is given by:
\begin{equation}
H = \frac{\omega_1 \sigma_{z}^1}{2} + \frac{\omega_2 \sigma_{z}^2}{2} + g(\sigma_{-}^1\sigma_{+}^2 + \sigma_{+}^1\sigma_{-}^2),
\end{equation}
where $\omega_1$ and $\omega_2$ are the qubit frequencies, $\sigma_{z}^i$ is the Pauli-z operator for qubit $i$, and $g$ represents the coupling strength between the qubits. The operators $\sigma_{+}^i$ and $\sigma_{-}^i$ are the raising and lowering operators for qubit $i$, respectively.
The qubits are operated near resonance, and the coupling strength $g$ is initially set to zero. We employ coherent driving terms $\sigma_{x}^1$ and $\sigma_{x}^2$, along with the tunable coupling strength $g$, as controls to manipulate the qubits. Each control pulse is discretized into 50 segments, allowing for fine-grained control over the system's evolution.
\begin{figure}
    \centering
    \includegraphics{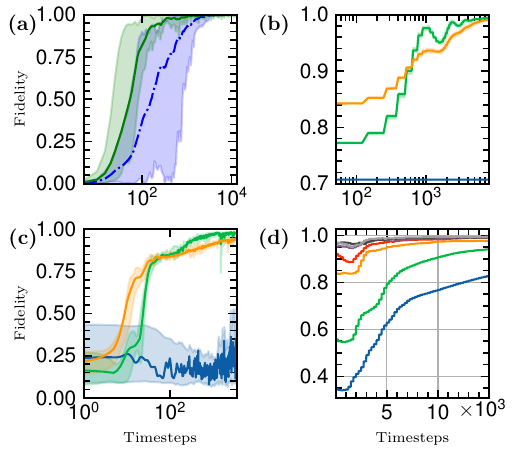}
    \caption{The performance of RLfD method.  (a) The variation in average reward throughout the training process of Bell state preparation. The blue line represents the performance of the original model-free RL method, and the green line corresponds to the RLfD method. The minimum number of timesteps required to achieve a fidelity of 0.995 is 1340 for the RLfD method, compared to 3060 for model-free RL.  (b) and (c) illustrate the fidelity achieved by the PPO and SAC algorithms, respectively, for preparing binomial states. The yellow and orange solid lines correspond to RLfD method with different initial fidelities, and the blue solid line corresponds to model-free methods. The shaded areas surrounding each line indicate the range of rewards, from best to worst, across various random seeds. (d) Fidelities with different filter bandwidths while preparing a cat state. The filters are arranged over a frequency range from $6.25\,\rm{MHz}$ to $81.25\,\rm{MHz}$ with a gradient of $12.5\,\rm{MHz}$, and they are color-coded as blue, green, orange, red, purple, brown and gray. The total bandwidth is set at $125\,\rm{MHz}$.}
    \label{fig:robust}
\end{figure}

Fig. \ref{fig:2}(a) and (d) shows the average reward throughout the training process. The original fidelity of pulses generated by GRAPE is around $0.9$. Because this task is relatively simple, both method and with different model bias have similar performance. The final infidelity is around $10^{-3}$ to $10^{-4}$ and the model can converge very quickly.

Fig. \ref{fig:robust}(a) shows the training stability of model-free RL and RLfD methods can convergence versus training timesteps. With the same number of training timesteps, the average reward achieved by the RLfD method is higher than that of the original model-free RL method. The shaded area of RLfD method is smaller, which indicates its robustness with different initial parameter setup.

\subsection{\label{sec:fock}Complex state preparation of bosonic system
}
\begin{figure}

    \includegraphics[]{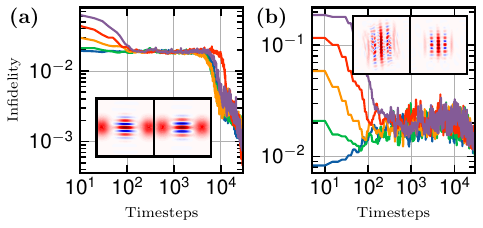}

    \caption{Quantum states preparation with RLfD method by optimizing  gate parameters . Different solid lines refer to different bias of gate parameters(5\%, 10\%, 15\%, 20\%, 25\% for blue, green, orange, red and purple solid lines respectively). 
        (a) Cat state preparation with gate depth of 5. (b) GKP state preparation with $\Delta = 0.3$. Gate depth is set to ten. The insets display the Wigner function of the state before training (left) and at the end of training (right) under a 25\% gate parameter error setup.
    }
    
    \label{fig:gate_code}
\end{figure}

Then we use RLfD method to prepare a superposition of Fock states or coherent states. Currently, quantum error correction (QEC) is being extensively studied in such states~\cite{ofek2016extending, ni2023beating}. Here, we use PLfD method to generate a binomial code state  given by $(\vert 0\rangle + \vert 4\rangle)/{\sqrt 2}$ and a even cat stae  $\vert \alpha \rangle + \vert -\alpha\rangle$ with $\alpha = 2$.
We consider a transmon qubit and a cavity in our physical model, which are dispersively coupled. With the consideration of high level leakage, qubit modes are considered as bosonic modes with kerr factor. The dimension of the Hilbert space is truncated to 3 and 7 for the qubit and cavity, respectively.
We use a Hamiltonian with consideration of Kerr factors~\cite{RevModPhys.93.025005},
\begin{equation}
    \begin{aligned}
    H & = \Delta \omega_c a^\dagger a + \Delta \omega_q b^\dagger b + \chi a^\dagger a b^\dagger b \\
    & \quad - \frac{E_c}{2} b^\dagger b^\dagger bb + \frac{K}{2} a^\dagger a^\dagger a a + \frac{\chi^\prime}{2} a^\dagger a^\dagger a a b^\dagger b.
\end{aligned}
\end{equation}
frequencies $\Delta \omega_c$ and $\Delta \omega_q$ are in the rotating frame of each driving frequency; $\chi$ cross Kerr factor between the qubit and the cavity; $E_c$ is the anharmonicity of the qubit; $K$ is the self Kerr factor of the cavity; and $\chi^\prime$ is the higher-order cross-Kerr factor.

The control Hamiltonians are complex drives of the qubit and resonator at their original frequencies. The experimental setup is the same as in ~ref.~\cite{heeres_implementing_2017}. There are four pulse shapes needed to be optimized, and each pulse is discretized into 275 segments over time, resulting in a total of 1100 parameters. We have also tried discretizing pulses to 550 segments in total of 2200 parameters. The model can also converge in this case.

Fig. \ref{fig:2}(b), (c), (e), and (f) illustrate how the infidelity varies with different initial fidelities and optimization methods. To simulate system imperfections, we also apply a sliding window filter and introduce up to 25\% of parameter fluctuation.

Comparing Fig. \ref{fig:2}(b) and (c) with (e) and (f), it is evident that regardless of the initial pulse infidelity, both methods lead to a reduction in infidelity. Notably, pulses with higher initial fidelity tend to converge faster and achieve a lower final infidelity in both the SAC and PPO methods.

In comparing the training algorithms, we observe that for the preparation of binomial code states and cat states, the PPO method outperforms in terms of both training stability and final infidelity. The enhanced stability is attributed to its on-policy nature, while the lower final infidelity results from the application of reward normalization.

We subsequently evaluated the robustness of both the SAC and PPO algorithms within the context of a binomial code state. As illustrated in Fig. \ref{fig:robust}(b), without the pre-training process, the PPO method is unable to address the high-dimensional quantum control problem when learning from scratch, as evidenced by the stagnation in the average reward during the training process. However, when demonstration data are introduced, the PPO algorithm converges reliably. The performance stability is observed across different initial random seeds, thereby indicating the robustness of randomization.

As shown in Fig. \ref{fig:robust}(c), the model-free SAC algorithm also fails to converge under a learning-from-scratch regime.  When demonstrations are provided, the SAC method achieves convergence. This result shows the critical role of demonstration data in improving the stability and convergence of reinforcement learning algorithms in high-dimensional quantum control tasks. 


PPO method demonstrates greater stability when pre-training is involved. The overall performance trend of the agent is upward, and no significant jitter is observed. The final reward of the PPO method is higher than that of the SAC method. The final fidelity of PPO method is about 99.8\% and SAC method is about 99\%. 

In Fig. \ref{fig:robust}(d), we illustrate the influence of channel bandwidth on state preparation. Our results indicate that when the bandwidth is sufficiently wide, the final fidelity remains high. However, when the bandwidth becomes overly constrained, the final fidelity reduces significantly. Notably, the RLfD method can mitigate this degradation by enhancing the final fidelity, although its performance does not fully match that achieved without bandwidth limitations.

We finally test the states generated by the gate parameters, showcasing cat states and Gottesman-Kitaev-Preskill (GKP) states prepared with echoed conditional displacement (ECD) gates~\cite{eickbusch2022fast, sivak2023real}, as illustrated in Fig. 4(a),(b). We consider the model bias as gate parameter errors here. We find that, regardless of the initial fidelity setup, the model converges to a relatively high fidelity that remains unchanged for an long period. After a long-time training, the model eventually achieves higher performance. This behavior may be due to the limited number of gate parameters, which makes it easy for the model to become trapped in local minima.




\section{\label{sec:discussion}CONCLUSIONS}
In this paper, we have introduced Reinforcement learning from demonstration (RLfD) into the field of quantum control, showcasing its ability to enhance control fidelity with the pulse shapes generated by methods like Gradient Ascent Pulse Engineering (GRAPE) or other sources. By integrating RLfD, we have demonstrated significant improvements in learning efficiency across various quantum control tasks.

Our analysis highlights the distinct features and advantages of both off-policy and on-policy reinforcement learning algorithms within the context of quantum control. Specifically, the off-policy algorithm, soft actor critic (SAC), although in some task it requires fewer interactions with the environment to reach a specified performance threshold and is capable of learning from scratch, the overall performance is not good enough while pursuing a very high fidelity. This limitation may be attributed to its off-policy nature; although SAC is efficient in exploring the environment, it is less adept at precisely determining specific pulse shapes.

Without demonstrations, PPO also struggles to handle quantum control problems with high-dimensional action spaces, as evidenced by its stagnant average reward during training. Nevertheless, when equipped with demonstrations and a pre-trained policy network, PPO excels. The prior knowledge significantly enhances its practicality, allowing it to converge to high-performance policies. The integration of demonstrations ensures greater training stability and convergence. This feature is crucial, especially in complex environments where prior knowledge can be extremely beneficial. Our findings suggest that PPO, when combined with RLfD, can be extended to more complex quantum control problems with high-dimensional action space.

In conclusion, we have shown that both simple and complex quantum control tasks can significantly benefit from RLfD. The samples required from interacting with the environment can be greatly reduced, thereby increasing the sample efficiency of reinforcement learning methods. For complex quantum environments with high-dimensional action space, RLfD assists agents in exploration and ultimately leads to convergence. The training process is also more stable, reducing variance across different runs and initial parameters.

Our findings open up an avenues for applying reinforcement learning techniques to quantum control, suggesting that incorporating demonstrations can overcome some of the inherent challenges posed by the high dimensionality and partial observability of quantum systems. Future work may explore the application of RLfD to even more complex quantum tasks, such as multi-qubit entanglement generation, error correction codes, and quantum algorithms, further bridging the gap between theoretical methods and practical quantum computing implementations.

\section{\label{Acknowledgement}ACKNOWLEDGMENTS
}

This work was supported the National Natural Science Foundation of China (NSFC) with Grant No. 62433015 and Laoshan Laboratory (LSKJ202200900). R.-B.W. was supported by National Natural Science Foundation of China with Grant No.62173201. J. Z. was supported by the Leading Scholar of Xi’an Jiaotong University and Innovative leading talent project "Shuangqian plan" in Jiangxi Province. P.S. was supported by "Young Talent Support Plan" of Xi'an Jiaotong University.


\appendix

\section{Reinforcement learning algorithm}
The communication time between the RL-agent and the Arbitrary Waveform Generator (AWG) is typically on the order of milliseconds. However, a single quantum control task, complete with reward measurement, can be executed in microseconds. This discrepancy allows for the acquisition of multiple trajectories $(s, a, r, s')$  within a single communication cycle with the AWG. Generally, smoother and differentiable rewards are good for stable training process. For non-feedback tasks, utilizing the average of multiple measurements as the reward is considered beneficial because these tasks typically involve a limited set of initial states, or even only one single initial state (state reset can be used to achieve this). This averaging approach helps stabilize the reward signal and enhances the reliability of the training process. 

The reinforcement learning from demonstration algorithms are based on the basic implementation of stable-baselines3~\cite{stable-baselines3}.

\begin{figure}

    \includegraphics[scale=1]{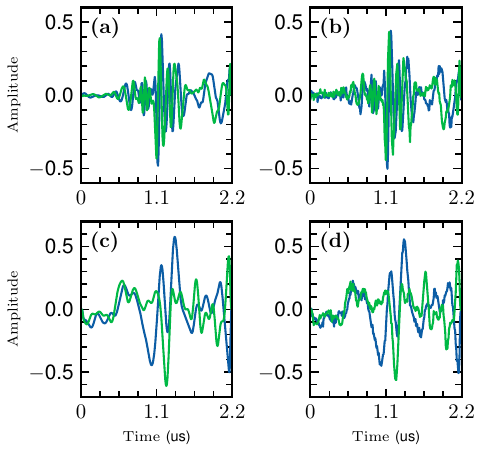}

    \caption{Comparison of pulse shapes. The blue line represents the in-phase component, while the green line represents the quadrature component.
        This figure shows the original demonstrations pulses and the pulses generated by RL agent after training. Demonstration of qubit driving (a) and cavity driving (c) before training.  Qubit driving (c) and cavity driving (d) generated by RL agent after training. }
    \label{fig:compare}
\end{figure}

\subsection{SAC algorithm}\label{app:sec:sac}
SAC maximizes the expectation of weighted average of reward and entropy of actions for better balance between exploration and exploitation. The definition of entropy of a distribution is:
\begin{equation}
H(P)=\underset{x\sim P}{\operatorname{E}}\left[-\log P(x)\right].
\end{equation}

$Q-$function is defined as 
\begin{equation}
\begin{aligned}
Q(s, a)=\underset{\tau \sim \pi}{\mathrm{E}}&\Big[\sum_{t=0}^{\infty} \gamma^{t} R\left(s_{t}, a_{t}\right)+ \\
& \alpha \sum_{t=1}^{\infty} \gamma^{t} H\left(\pi\left(\cdot \mid s_{t}\right)\right) \mid s_{0}=s, a_{0}=a\Big],
\end{aligned}
\end{equation}
the optimal policy is
\begin{equation}
\pi^*=\arg\max_\pi\underset{\tau\sim\pi}{\operatorname{E}}\left[\sum_{t=0}^\infty\gamma^t\left(R(s_t,a_t)+\alpha H\left(\pi(\cdot|s_t)\right)\right)\right].
\end{equation}

Intuitively, the entropy term is used to prevent the variance of action from going to zero, thereby keep randomness of exploration.

The loss of $Q$-function is defined as 

\begin{equation}
J_{Q}(\phi)=\underset{\tau \sim \mathcal{D}}{\mathrm{E}}\left[\frac{1}{2}\left(Q_{\phi}\left(s_{t}, a_{t}\right)-\left(R\left(s_{t}, a_{t}\right)+\gamma V_{\phi, \rm{min}}(s^\prime)\right)\right)^{2}\right],
\end{equation}

where 
$$V_{\phi, \rm{min}} = Q_{\phi, \min }\left(s_{t+1}, a_{t+1}\right)-\alpha \log \pi\left(a_{t+1} \mid s_{t+1}\right), $$ where $Q_{\phi,\rm{min}}(s, a) = \min\left(Q_{\phi_1}(s, a), \ Q_{\phi_2}(s, a)\right) $.

The $J_{\pi, \rm{SAC}}$ is defined as below
\begin{equation}
J_{\pi, \rm{SAC}}(\theta)
= \underset{{\tau \sim \mathcal{D}}, \ a \sim \pi_\theta}{\mathrm{E}} \Bigg[ \log \pi_{\theta}\left(a_{t} \mid s_{t}\right) - \frac{1}{\alpha} Q_{\phi,\rm{min}}\left(s_{t}, a_{t}\right) \Bigg].
\end{equation}
In practice, this loss is always implemented as a surrogate loss with the help of reparameterization trick.

The algorithm SACfD for Quantum Control is shown below.

\begin{algorithm}[H]
    \caption{SACfD for Quantum Control}\label{alg:sac}
    \begin{algorithmic}
    \Require
    \State $H_0$: Drift Hamiltonian, $H_k$: Control Hamiltonian
    \State $u_0$: Initial guess of discrete control pulses
    \State $T$: Total time, $\Delta t$: Time step size
    \State $\{\vert \psi_k\rangle \}$ possible initial state set
    \State $\vert C\rangle $ target state
    \State $\theta$: Initial policy parameters
    \State $\phi_1, \phi_2$: $Q-$function parameters
    \State $\mathcal{D}$: Replay buffer
    \State $\mathcal{M}$: Demonstration buffer
    \State $\mu$: Ratio of demonstration data in a mini-batch
    
    \Ensure Trained parameters $\theta$, $\phi_1$, $\phi_2$
    
    \State $\mathcal{D} \gets \{\}$, $\mathcal{M} \gets \{\}$
    
    \For {$\vert \psi_k\rangle \ \texttt{in}\  \{\vert \psi_k\rangle \}$ }
        \State $u^*_k \gets \text{GRAPE}(\vert \psi_k\rangle, \vert C\rangle, u_0, H_0, H_k, T, \Delta t)$ 
        \State $a_k \gets u^*_k$, $s_k \gets \vert \psi_k\rangle$
        \State Interact with environment to get $(s_k, a_k, r, s', d)$
        \State $\mathcal{M} \gets \mathcal{M} \cup \{(s_k, a_k, r, s', d)\}$ \Comment{Store data to $\mathcal{M}$}
    \EndFor
    \State Construct deterministic expert policy $\pi^*(s) \to a$ with $\mathcal{M}$ 
    
    \For {$e\  \texttt{epochs}$}
    \For {$i\ \texttt{in range(k)}$}
    \State sample data from environment $(s, a, r, s', d)$
    \State $\mathcal{D} \gets \mathcal{D} \cup \{(s, a, r, s', d)\}$
    \EndFor
    \State Sample a mini-batch $B$ from $\mathcal{D}$ and $\mathcal{M}$ according to $\mu$. 
    
    \State Calculate $\nabla J_Q$ and update $\phi_1, \phi_2$ with $\nabla J_Q$
    \State Calculate $\nabla J_\pi$ with $$
        J_\pi=\frac{1}{|B|} \sum_{(s, a) \in B}\left(J_{\pi, \rm{SAC}} + \lambda_{\rm{BC}}\|\pi_e(s) - \pi^*(s)\|_2^2\right)$$
    \State Update $\theta$ with $\nabla J_\pi$
    
    \EndFor
    
\end{algorithmic}
\end{algorithm}

The behavior cloning coefficient $\lambda_{\rm{BC}}$ is set to 2 according to ~\cite{martin2021learning}. And we add an exponential decay to $\lambda_{\rm{BC}}$. This will let the agent learn more from demonstrations in the early stage of training and learn more from the environment after the initial stage. In some tasks, behavior cloning loss may be considered unnecessary~\cite{ball2023efficient}. In complex quantum control problems, because the action space dimension is rather high, behavior cloning loss can provide significant performance guarantees.

\subsection{PPO algorithm}\label{app:sec:ppo}
PPO uses clip trick to prevent updates from being large, thereby reducing the possibility that the performance drops after updates. Basic updates follow

\begin{equation}
    \theta_{k+1}=\arg \max _{\theta} \underset{s, a \sim \pi_{\theta_{k}}}{\mathrm{E}}\left[L\left(s, a, \theta_{k}, \theta\right)\right],
\end{equation}

where

\begin{widetext}
    \begin{equation}
    L\left(s, a, \theta_{k}, \theta\right)=\min \left(\frac{\pi_{\theta}(a \mid s)}{\pi_{\theta_{k}}(a \mid s)} A^{\pi_{\theta_{k}}}(s, a), \quad \operatorname{clip}\left(\frac{\pi_{\theta}(a \mid s)}{\pi_{\theta_{k}}(a \mid s)}, 1-\epsilon, 1+\epsilon\right) A^{\pi_{\theta_{k}}}(s, a)\right).
\end{equation}
\end{widetext}

where $\epsilon$ is a (small) hyperparameter which refers to how much the new policy is allowed to deviate from the old policy. $A^{\pi_{\theta_{k}}}(s, a)$ is the advantage function defined as 
\begin{equation}
    A(s, a)=R+\gamma \cdot V_\phi(s^{\prime})-V_\phi(s),
\end{equation}
where $s^\prime$ is the next state after $s$ in the trajectory.

The algorithm PPO with pre-train for Quantum Control is shown in Algorithm 2.

$H(\cdot|s_t)$ is the entropy of $\pi_\theta(\cdot|s_t)$, it’s used to maintain randomness of $\pi_\theta(\cdot|s_t)$ to preserve the exploration ability. Entropy here is used to prevent the policy’s variance from being zero. Here $\lambda_{\rm{ent}}$ is the coefficient of entropy that controls the extent of exploration by influencing how much randomness is introduced into the policy.

We finally show the comparison of pulses shapes generated by RL agent and original demonstrations in Fig \ref{fig:compare}. We can see that the pulse shapes generated by the RL agent are only slightly different from the original demonstrations. These subtle differences indicate that the agent is capable of fine-tuning the pulse shapes to better adapt to the environment. By making minor adjustments, the RL agent can compensate for system imperfections or calibration errors, leading to improved performance in the quantum control task.

\begin{algorithm}[H]
    \caption{PPO with pre-train for Quantum Control}\label{alg:ppo}
    \begin{algorithmic}
    \Require
    \State $H_0$: Drift Hamiltonian, $H_k$: Control Hamiltonian
    \State $u_0$: Initial guess of discrete control pulses
    \State $T$: Total time, $\Delta t$: Time step size
    \State $\{\vert \psi_k\rangle \}$ possible initial state set
    \State $\vert C\rangle $ target state
    \State $\theta$: Initial policy parameters
    \State $\phi$: $V-$function parameters
    \State $\mathcal{D}$: Rollout buffer
    \State $\mathcal{M}$: Demonstration buffer
    \State $\mu$: Ratio of demonstration data in a mini-batch
    
    \Ensure Trained parameters $\theta$, $\phi_1$, $\phi_2$
    
    \State $\mathcal{D} \gets \{\}$
    \State $\mathcal{M} \gets \{\}$
    
    \For {$\vert \psi_k\rangle \ \texttt{in}\  \{\vert \psi_k\rangle \}$ }
        \State $u^*_k \gets \text{GRAPE}(\vert \psi_k\rangle, \vert C\rangle, u_0, H_0, H_k, T, \Delta t)$ 
        \State $a_k \gets u^*_k$, $s_k \gets \vert \psi_k\rangle$
         \State Interact with environment to get $(s_k, a_k, r, s', d)$
        \State $\mathcal{M} \gets \mathcal{M} \cup \{(s_k, a_k, r, s', d)\}$ \Comment{Store data to $\mathcal{M}$}
    \EndFor

    \For {$i\ \texttt{in range(k)}$}
    \State Sample a set of trajectory $D=\{\tau_i\}$ from demonstration buffer $\mathcal{M}$  
    
    \State Calculate $\pi_\theta(a_t|s_t), V_\phi(s_t), H(\cdot|s_t)$ with $\theta,\phi$
    \State Update $\phi_1, \phi_2$ with $$
    J = \frac{1}{|D|T}\sum_{\tau_i \in D}\sum^T_{t=0}\Big(\nabla\log\pi_\theta(a_t|s_t)+ \lambda_{\rm{ent}}H(\cdot|s_t)\Big )
    $$
    \State Updating $V-$function by minimizing $$L=\frac{1}{\left|D\right| T} \sum_{\tau_i \in {D}} \sum_{t=0}^{T}\left(V_{\phi}\left(s_{t}\right)-\hat{R}_{t}\right)^{2}$$
\EndFor

\State Using the normal PPO method with pre-trained agent parameters with quantum systems(environments).

\end{algorithmic}
\end{algorithm}

\nocite{*}

\bibliography{ref}

\end{document}